\documentclass{article}
\usepackage{spconf,amsmath,graphicx}

\title{AclNet: efficient end-to-end audio classification CNN}
\name{Jonathan J Huang, Juan Jose Alvarado Leanos}
\address{Intel Corporation}

\begin{document}
%
\maketitle
\begin{abstract}

We propose an efficient end-to-end convolutional neural network architecture, AclNet, for audio classification.  When trained with our data augmentation and regularization, we achieved state-of-the-art performance on the ESC-50 corpus with $85.65\%$ accuracy.  Our network allows configurations such that memory and compute requirements are drastically reduced, and a tradeoff analysis of accuracy and complexity is presented. The analysis shows high accuracy at significantly reduced computational complexity compared to existing solutions.  For example, a configuration with only $155k$ parameters and $49.3$ million multiply-adds per second is $81.75\%$, exceeding human accuracy of $81.3\%$.  This improved efficiency can enable always-on inference in energy-efficient platforms.  

\end{abstract}
\begin{keywords}

Convolutional neural networks, end-to-end CNN, environmental sound classification, audio classification

\end{keywords}
\section{Introduction}
\label{sec:intro}

Following the successes of image classification, convolutional neural network (CNN) architectures have become popularized for audio classification.  Hershey, et al. \cite{hershey2017cnn} showed that large image classification CNNs trained with huge amount of weakly labeled Youtube data leads to semantically meaningful representations, the basis of a powerful classifier.  In the recent DCASE acoustic scene classification task \cite{Mesaros2018_DCASE}, the top submissions are mostly CNN-based \cite{Sakashita2018} \cite{Dorfer2018}, \cite{Zeinali2018}.  Likewise, many of the top results for the ESC-50 corpus \cite{piczak2015esc} use various forms of CNNs \cite{sailor2017unsupervised}, \cite{tokozume2018learning}, \cite{kumar2018knowledge}, \cite{tak2017novel}.  

While the prior work on CNN audio classifiers have focused on accuracy of the performance for a particular tasks, none that we are aware of have treated computational efficiency as a primary objective.  The first contribution of this paper is a scalable architecture, that at the high-end has one of the best accuracies in the ESC-50, and at the low-end offers the flexibility to scale down to extremely small model sizes.  The advantage of a scalable architecture is that it allows for flexibility in inference platforms with various system constraints.  The efficiency brought by our architecture allows for low-power always-on inference on DSPs or dedicated neural net accelerators \cite{deisher2017gna}, \cite{ionica2015movidius}.  The second contribution is the application of mixup data augmentation \cite{zhang2017mixup} for audio classification, and we show it is a big contributor to the high accuracy due to improved generalization.

AclNet is an end-to-end (e2e) CNN architecture, which takes raw time-domain input waveform as opposed to the more popular technique of using spectral features like mel-filterbank or mel-frequency cepstral coefficients (MFCC).  One of the advantages of an e2e architecture is that the front-end feature makes no assumptions of the spectral content.  Its feature representation is learned in a data-driven manner, thus its features are optimized for the task at hand as long as there is sufficient training data.  Another advantage of e2e is that it eliminates the implementation of the spectral features, which simplifies software or hardware complexity.  Although other e2e techniques \cite{aytar2016soundnet}, \cite{Dai2017VeryDC}, \cite{tokozume2017learning}, \cite{tokozume2018learning} have been studied, our architecture has a focus on efficiency.  

Several research results from optimization of CNNs from image domain can be borrowed to make audio classification more efficient.  Han et al. \cite{han2015deep} used pruning, quantization, and Huffman encoding to reduce complexity of CNNs.  Singular value decomposition has been applied to DNN to reduce model size \cite{xue2014singular}.  The AclNet gets its inspirations for efficient computations from MobileNet \cite{howard2017mobilenets}, which features the depthwise separable convolution we employed extensively in this work.  With these tricks, human-level accuracy for ESC-50 was achieved with only $155k$ parameters and $49.3$ million multiply-adds per second (MMACS).  

The remainder of this paper is organized as follows.  In Section \ref{sec:aclnet} we detail the AclNet architecture.  Section \ref{sec:experiment} provides the details of data augmentation and model training process.  Section \ref{sec:results} presents the our findings from the experiments, followed by conclusion in Section \ref{conclusion}.

\section{AclNet architecture}
\label{sec:aclnet}

The AclNet architecture consists of two building blocks of the network:  the low-level features (LLF) shown in Table \ref{table_llf} and the high-level features (HLF) shown in Table \ref{table_hlf}.

\subsection{Low-level features}
\label{ssec:llf}

The LLF can be viewed as a replacement of the spectral feature, and the two stages of 1-D strided convolutions are equivalent to FIR decimation filterbank.  With the time-domain waveform as input, the LLF produces an output of 64 channels at feature frame rate of $10 ms$ after the maxpool layer.  In the example given on Table \ref{table_llf}, the $1.28 s$ input produces an output tensor with dimension $(64,1,128)$.

Although the number of parameters in the LLF are invariant to stride values $S1$, $S2$ and the number of intermediate channels $C1$, the choice of these values greatly influence the compute complexity and accuracy.  Our experiments will show the settings which gives the most accurate results. 

The example in Table \ref{table_llf} is for the $16kHz$ sampling rate.  The parameters $S1$, $S2$, and all kernel size scale linearly with sampling rate to ensure the same time duration of kernels and output frame rate of $10 ms$.

\hbadness=99999
\begin{table}[h]
	\caption{AclNet low-level features, with $1.28$s $16 kHz$ samples.  Input dimension $(1,1,20480)$.} \label{table_llf}
\begin{center}
	\begin{tabular}{  p{1.2cm}  p{0.7cm}  p{2.7cm} p{0.6cm} p{1.3cm}}
		\hline\hline
Layer&Stride&Out dim&Out Chans&Kernel size \\ \hline
Conv1&S1&$C1,1,20480/S1$&C1&9\\
Conv2&S2&$64,1,20480/(S1 S2)$&64&5\\
Maxpool1&1&$64,1,128$&64&$160/(S1 S2)$\\
	\end{tabular}
\end{center}
\end{table}

\subsection{High-level features}
\label{ssec:hlf}

Continuing from the LLF output, transposing the first two dimensions will result in an image-like tensor with dimension $(1,64,128)$.  The rest of HLF thus can follow the structure similar to image classification CNNs.  We experimented with a number of architectures, and found that a VGG-like architecture \cite{simonyan2014very} provides a good classification performance and well-understood building blocks.  The architecture shown in Table \ref{table_hlf} is a modified VGG.  Besides changing the depth and channel width, the final layers of the network are also modified.  Conv12 layer is a $1\times1$ convolution that reduces the number of channels to the number of classes, which in the case of ESC-50, is $50$.  Each of the $50$ channels are then average pooled over the $2\times4$ patches and output as softmax.  The advantage of these final two layers is that our architecture can incorporate arbitrary length inputs, without any need to modify the number of hidden units in fully-connected layers.  An additional benefit of this way of pooling is that it is shown to be effective for training on weakly labeled datasets \cite{Shah2018ACL}.

Before the input to Conv12 layer, we have a dropout layer.  We found the probability of 0.2 to work well on this dataset.

\begin{table}[h]
	\caption{AclNet high-level features, with input $dimension=(1,64,128)$ out of LLF.} \label{table_hlf}
\begin{center}
	\begin{tabular}{  p{1.5cm}   p{0.7cm}  p{1.9cm} p{0.7cm} p{1.7cm}}
		\hline\hline
Layer&Stride&Out dim&Out Chans&Kernel Size \\ \hline
Conv3&1&$32,64,128$&32&$3\times3$\\
Maxpool2&1&$32,32,64$&32&$2\times2$\\ \hline
Conv4&1&$64,32,64$&64&$3\times3$\\
Conv5&1&$64,32,64$&64&$3\times3$\\
Maxpool3&1&$64,16,32$&64&$2\times2$\\ \hline
Conv6&1&$128,16,32$&128&$3\times3$\\
Conv7&1&$128,16,32$&128&$3\times3$\\
Maxpool4&1&$128,8,16$&128&$2\times2$\\ \hline
Conv8&1&$256,8,16$&256&$3\times3$\\
Conv9&1&$256,8,16$&256&$3\times3$\\
Maxpool5&1&$256,4,8$&256&$2\times2$\\ \hline
Conv10&1&$512,4,8$&512&$3\times3$\\
Conv11&1&$512,4,8$&512&$3\times3$\\
Maxpool6&1&$512,2,4$&512&$2\times2$\\ \hline
Conv12&1&$50,2,4$&50&$1\times1$\\
Avgpool1&1&$50$&50&$2\times4$\\
		
	\end{tabular}
\end{center}
\end{table}

\subsection{Convolutional layers details}
\label{ssec:conv_layers}
All convolutional layers shown on Tables \ref{table_llf} and \ref{table_hlf}, except their first layer (i.e. Conv1 and Conv3) can be configured in one of two forms:

\begin{itemize}
	\setlength\itemsep{0em}
	\item \textbf{Standard convolution (SC)}: this is the standard building blocks of convolution layer, batch normalization, and ReLU activation.

	\item \textbf{Depthwise separable convolution (DWSC)}: the convolution is decomposed into depthwise separable convolutions with pointwise layers each followed by batch normalization and ReLU activation as in MobileNet \cite{howard2017mobilenets}.
	
\end{itemize}

The advantage of DWSC is that they use significantly less parameters and operations compared to SC, but typically at a cost of degradation in performance.  We will explore the tradeoffs between these two choices of convolutions in our experiments.

\subsection{Width multiplier}
\label{ssec:width_mult}

As in MobileNet, our architecture also has a width multiplier (WM) to control the complexity of the network.  The WM linearly scales the number of output channels from Conv3 to Conv11.  This parameter is an easy way to manage the capacity of the network, and our experiments will explore its accuracy impact on the ESC-50 corpus.

\section{Experimental methods}
\label{sec:experiment}

\subsection{Dataset}
We used ESC-50 to train and evaluate the models. ESC-50 contains a total of 2000 examples of environmental sounds arranged in 50 classes. We use the default 5-folds provided by the dataset for cross validation in performance evaluation.  All sound files were converted to 16-bits, at $16 kHz$ and $44.1 kHz$ sampling rates for two different sets of experiments.  We eliminated the silent sections at the beginning and ending of each recording. 

\subsection{Data augmentation}
\label{ssec:augmentation}

We have experimented with different input lengths to the training of AclNet using ESC-50 and a proprietary dataset.  Empirically we found that between 1 to 2 second input gave the best results, so for the rest of the experiments we chose $1.5 s$ input length.  In our data loader of the training process, we use the following real-time data augmentation to generate each training example.  

\begin{enumerate}
	\setlength{\itemsep}{-0.5pt}
	\item Choose a random $2 s$ of audio within a training file
	\item Center the waveform to zero mean, and normalize by standard deviation
	\item Resample the waveform by a random factor uniformly chosen in range $[0.8, 1.25]$
	\item Crop exactly to $1.5 s$
	\item Multiply waveform by random gain chosen uniformly in range $[-6.0, +6.0] dB$ 
\end{enumerate}

During test time, only the data normalization step is used, and the length of the entire wave file is input into the network.

\subsection{Mixup training}
\label{ssec:mixup}

Mixup \cite{zhang2017mixup} is a recent technique to improve generalization by increasing the support of the training distribution. In this technique, a neighborhood is defined around each example in the training data by constructing virtual training examples, that is, pairs of virtual samples and virtual targets $\tilde{x},\tilde{y}$.
Given two training examples, $(x_i,y_i)$ and $(x_j,y_j)$, the new virtual pair is computed as:
\begin{equation} \label{eq:1}
	\tilde{x} = \lambda x_i + (1-\lambda) x_j
\end{equation}
\begin{equation} \label{eq:2}
	\tilde{y} = \lambda y_i + (1-\lambda) y_j
\end{equation}
			
where $\lambda \sim beta(\alpha,\alpha)$.  The hyperparameter $\alpha$ controls the amount that is mixed in from the second example. Higher values of alpha make the virtual pairs less similar to the original unmixed training examples.  We experimented with values from 0.1 to 0.5.

\subsection{Learning Settings}
For all experiments, we used stochastic gradient descent optimizer with momentum of $0.9$, weight decay of $2\textrm{e-4}$, and a batch size of $64$.  We trained the model using the following learning rate schedule with 3 different phases: 0.2 for the first 500 epochs, 0.04 for the next 1000, and 0.016 for the last 500, for a total of 2000 epochs for each fold. Also, for the first 100 epochs we disabled the mixup procedure as a form of warm up to improve initial convergence.

\section{Results and analysis}
\label{sec:results}

\subsection{Data augmentation and mixup}
\label{ssec:results_augmenentation}

\begin{figure}
	\includegraphics[width=\linewidth]{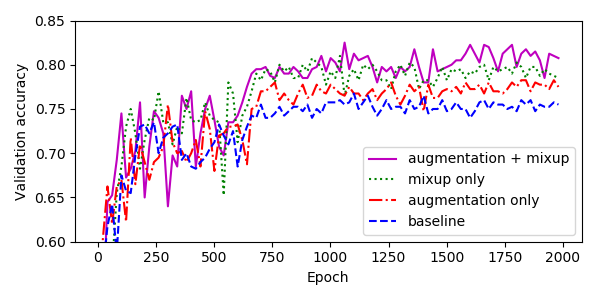}
	\caption{Comparison of the effects of data augmentation and mixup on validation accuracy.}
	\label{fig:val_acc}
\end{figure}

Several experiments were done to assess the effectiveness of augmentation and mixup.  Figure \ref{fig:val_acc} shows the validation accuracy over the course of the training process for various combinations of augmentation as explained in Section \ref{sec:experiment}.  All experiments were done using a WM of 1.0, and SC.  We see an obvious improvement with each individual augmentation, and that mixup by itself is more effective than the other form of augmentation.  The best result was achieved when augmentation was combined with mixup, which had an absolute improvement of more than $5\%$ above the baseline without any augmentation.  We note mixup is conceptually similar to between class learning, which was also shown to work well for ESC-50 \cite{tokozume2018learning}.

We have experimented with the choice of $\alpha$ in mixup, and found that values between $0.1$ to $0.2$ worked well for the larger size architectures, thus for the remainder of experiments, we default to using this combined augmentation with mixup $\alpha=0.1$.

\begin{table*}[t]
	\caption{ESC-50 5-fold accuracies with AclNet at select configurations.}
	\bigskip
	\centering

	\begin{tabular}{  | p{1.3cm} | p{1.0cm} | p{1.3cm} | p{1.5cm} | p{1.3cm} | p{1.5cm} | p{1.3cm} | p{1.5cm} | p{1.5cm} | p{1.5cm} |} \hline\hline
		Sampling rate& Conv type &LLF params (k) &LLF MMACS &HLF params (k) &HLF MMACS &Total params (k) &Total MMACS & Width multiplier &Accuracy (\%) \\ \hline\hline
		16k & DWSC & 1.44 & 4.35 & 13.91 & 2.93 & 15.35 & 7.28 & 0.125 & 75.38 \\ \hline
		16k & DWSC & 1.44 & 4.35 & 153.43 & 31.07 & 154.87 & 35.42 & 0.5 & 80.40 \\ \hline
		16k & DWSC & 1.44 & 4.35 & 567.92 & 113.7 & 569.4 & 118.1 & 1.0 & 80.90 \\ \hline
		44.1k & DWSC & 1.81 & 17.98 & 13.91 & 2.96 & 15.72 & 20.94 & 0.125 & 75.50 \\ \hline
		44.1k & DWSC & 1.81 & 17.98 & 153.43 & 31.33 & 155.23 & 49.31 & 0.5 & 81.75 \\ \hline
		44.1k & DWSC & 1.81 & 17.98 & 567.92 & 114.6 & 569.73 & 132.59 & 1.0 & 83.10 \\ \hline
		44.1k & SC & 6.99 & 80.9 & 77.21 & 8.88 & 84.21 & 131.17 & 0.125 & 82.30 \\ \hline
		44.1k & SC & 6.99 & 80.9 & 1190.0 & 132.72 & 1197.0 & 255.01 & 0.5 & 83.95 \\ \hline
		44.1k & SC & 6.99 & 80.9 & 4730.0 & 524.67 & 4737.0 & 646.97 & 1.0 & 85.0 \\ \hline
		44.1k & SC & 6.99 & 80.9 & 10620 & 786.56 & 10627 & 867.45 & 1.5 & \textbf{85.65} \\ \hline

	\end{tabular}
	\label{table_accuracy}

  \end{table*}

\subsection{Low-level feature parameters}
\label{ssec:llf_params}

In EnvNet \cite{tokozume2017learning}, analysis showed that 2 convolutions of kernel size 8 worked best for this dataset.  Our experiments confirmed that 2 convolutions being optimal, but we also found that slightly reducing the kernel size of second convolution had no impact on accuracy.  Our best setting is with kernel sizes of 9 and 5 for the first two convolutions.

In order to determine the choice of other LLF parameters, we did a grid search of the parameter space over these ranges: $C1 \in \{8, 16, 32\}$, $S1 \in \{2, 4, 8\}$, $S2 \in \{2,4\}$.

We trained AclNet using both SC and DWSC settings with width multiplier of 1.0, and found the values of (C1, S1, S2) = $(8, 2, 2)$ for SC and $(16, 2, 4)$ for DWSC gave the best accuracy.  For the remainder of experiments, we will default to using these best settings for SC and DWSC.  The experiments showed that there was about a $3\%$ difference between the best and worst parameters for each of the settings.  The best result in both cases was not the highest complexity, which is $(32, 2, 2)$.  We suspect the heavier LLF settings might be overfitting, and that with more training data we could reach a different conclusion.

\subsection{Complexity versus accuracy}
\label{ssec:accuracy_analysis}

\begin{figure}
	\includegraphics[width=\linewidth]{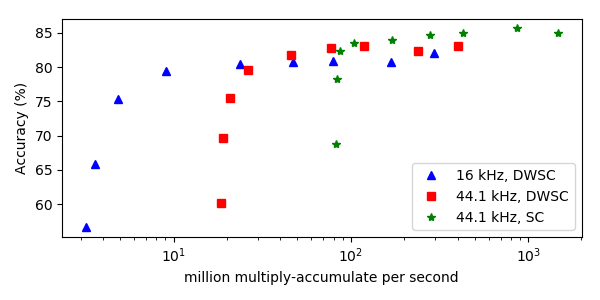}
	\caption{Accuracy vs million multiply-adds per second.}
	\label{fig:acc_mmacs}
\end{figure}

To understand the tradeoff between complexity and accuracy, we ran three sets of experiments using 1) $16 kHz$ input with DWSC, 2) $44.1 kHz$ input with DWSC, and 3) $44.1 kHz$ input with SC.  For each set, we did the 5-fold validation with WM configured at $1/32$, $1/16$, $0.125$, $0.25$, $0.5$, $0.75$, $1.0$, $1.5$, and $2.0$.  Figure \ref{fig:acc_mmacs} shows the accuracy versus MMACS for each of the settings, color-coded by sets.  For each of these settings, increasing complexity generally led to better accuracy.  The exception is at the highest WM, where it is possible that we hit diminishing returns of higher capacity.  In all cases, WM below $0.25$ steepens the drop in accuracy.  Another observation is that for the same HLF settings, $44.1 kHz$  sampling rate improves accuracy by around $2\%$.   

Table \ref{table_accuracy} shows a subset of these experiments, with details of LLF, HLF, overall complexity and accuracy. Our best accuracy of $85.65\%$ was achieved with $44.1 kHz$  sampling rate, SC, and $1.5$ WM.  At the time of this writing, this is the best single system accuracy reported for ESC-50 (second  overall behind an ensemble system \cite{sailor2017unsupervised}).  With DWSC models, we can see that the total parameter and MMACS are significantly lower than SC for the same WM.  The result on $44.1$ kHz, DWSC, and $0.5$ WM has $81.75\%$, which exceeds human accuracy of $81.3\%$ \cite{piczak2015esc}, was achieved with only $155k$  parameters, and $49.31$ MMACS.  We note that human accuracy is also exceeded with SC, WM of 0.125, a model that has a modest $84k$ parameters and $131.17$ MMACS.  As a comparison of complexity, EnvNetV2 \cite{tokozume2018learning}, which at the time of this writing has the best single model accuracy of $84.9\%$, uses $101M$ parameters and $1033$ MMACS.  Our best model with accuracy of $85.65\%$ has about $1/10$ the parameters and $16\%$ less operations.

\section{Conclusion}
\label{conclusion}

We have presented a novel e2e CNN architecture, AclNet, for audio classification.  AclNet is a scalable architecture that achieved state-of-the-art $85.65\%$ accuracy with high compute, and better than human level accuracy of $81.75\%$ with only $155k$ parameters and $49.3$ MMACS.  To achieve the low complexity with high accuracy, AclNet used depthwise separable convolution blocks.  The combination of mixup and data augmentation helped boost the accuracy by $5\%$, which had a major contribution to achieving one of the best results reported on ESC-50 dataset.  

\bibliographystyle{IEEEbib}
\bibliography{refs}

\end{document}